\documentclass[aps,prd,twocolumn,superscriptaddress]{revtex4}
\usepackage{epsfig,epsf}
\usepackage{amsmath}
\usepackage{amsthm}
\usepackage{amsfonts}
\usepackage{amssymb}
\usepackage{dsfont}
\usepackage{multirow}
\usepackage{appendix}
\usepackage{slashed}
\usepackage[active]{srcltx}
\usepackage{psfrag}
\usepackage{subfigure}

\begin{document}

\title{Light scalar  $K_{0}^{*}(700)$ meson in vacuum and  a hot medium}
\date{\today}

\author{K.~Azizi}
\thanks{Corresponding author}
\affiliation{Department of Physics, University of Tehran, North Karegar Avenue, Tehran
14395-547, Iran}
\affiliation{Department of Physics, Do\v{g}u\c{s} University, Acibadem-Kadik\"{o}y, 34722
Istanbul, Turkey}
\author{B.~Barsbay}
\affiliation{Department of Physics, Do\v{g}u\c{s} University, Acibadem-Kadik\"{o}y, 34722
Istanbul, Turkey}
\affiliation{Department of Physics, Kocaeli University, 41380 Izmit, Turkey}
\author{H.~Sundu}
\affiliation{Department of Physics, Kocaeli University, 41380 Izmit, Turkey}

\begin{abstract}
The $K_{0}^{*}(700)$ meson appears as the lightest strange scalar meson in PDG. Although there were a lot of experimental and theoretical efforts to establish this   particle and determine its properties and nature, it still needs confirmation in an experiment and its internal quark-gluon organization needs to be clarified. In this connection, we study some spectroscopic properties of this state in a hot medium as well as a vacuum by modeling it as a usual meson of a quark and an aniquark. In particular, we investigate its mass and coupling or decay constant in terms of the temperature of a hot medium by including the medium effects by the fermionic and gluonic parts of the energy momentum tensor as well as the  temperature-dependent continuum threshold, quark, gluon and mixed condensates. We observe that the mass of $K_{0}^{*}(700)$ remains unchanged up to $T \simeq 0.6 ~ T_c$ with $ T_c $ being the critical temperature, but it starts to diminish after this point and approaches zero near to the critical temperature referring to the melting of the meson. The coupling of $K_{0}^{*}(700)$ is also sensitive to $ T $ at higher temperatures. It  starts to grow rapidly after $T \simeq 0.85 ~ T_c$. We turn off the medium effects and calculate the mass and coupling of   the $K_{0}^{*}(700)$ state at zero temperature.  The obtained mass is in accord with the average Breit-Wigner mass value reported by PDG.
\end{abstract}

\maketitle

\section{Motivation}
The light scalar mesons with a mass below $ 1~\mathrm{GeV} $ are among particles that are not experimentally well established and their nature needs to be clarified. Their mass and width suffer from large uncertainties. Hence, their investigation constitutes one of the directions of research in high energy physics. There are a lot of models and approaches aiming to clarify their nature and internal quark-gluon organization. The standard  quark model for the mesons 
handling  them as bound states of quarks and antiquarks  fails to correctly describe the mass hierarchy and the existing large uncertainties on the parameters of 
these particles. As a result, some of these particles like the light unflavored  $f_0(500)  $  state  have already been treated as the 
unconventional exotic states of compact tetraquarks or two-meson molecules \cite{Jaffe,Weinstein}.

The lightest scalar strange meson $K_{0}^{*}(700)$ appears in PDG with the quantum numbers $ I(J^P)=1/2(0^+) $  and the  qualifier ``needs confirmation". The reported average T-matrix pole as well as  Breit-Wigner mass and width  for this state are \cite{PDG}:
\begin{eqnarray}
&&\mbox{T-matrix pole}~\sqrt{s}   = (630-730)-i (260-340)~MeV,\nonumber\\
&&\mbox{Breit-Wigner mass}   = 824\pm 30~MeV,\nonumber\\
&&\mbox{Breit-Wigner width}   = 478\pm 50~MeV.
\end{eqnarray}
Its other name is $ \kappa $, and it has appeared in previous versions of PDG as  $K_{0}^{*}(800)$. With the uncertain  mass and width as well as large  width value and the fact that it  resides close to the $ K\pi $ threshold  and appears as a ``shoulder'' of the $K^*(892)$ in this invariant mass distribution make its establishment more difficult, experimentally. The BES-II Collaboration was found a $K_{0}^{*}(700)$-like structure in 2006 in the process $ J/\psi \rightarrow \bar{K}^{*0} K^{+}\pi^{-}$ \cite{BES-II}. The Belle Collaboration studied this state in the process $ \tau^{-}\rightarrow K_{S}^{0} \pi^{-}\nu_{\tau}$ \cite{Belle}. Many phenomenological approaches have been tried to explain this state and clarify the situation with it in vacuum (see, for instance, Refs. \cite{Cawlfield,Anisovich,Delbourgo,Oller,Shakin,Scadron,Bugg,Zheng,Zhou,Link,Aubert,Kopp,Jamin,Black,Genon,Pelaez,Humanic:2018brf}). Mainly, this state is considered as the usual strange scalar meson of $ d\bar{s} $, however, the  isospin, mass and decay channels of $K^*_0(700)$ state fit well in the  tetraquark nonet with low-mass that was predicted~\cite{Alford:2000mm}. But, no direct and powerful experimental sign for it to have a tetraquark  structure  exists \cite{Humanic:2018brf}. 
Investigation of the scalar meson $K_{0}^{*}(800)$ by modeling it as a scalar tetraquark of  diquark-antidiquark structure was made in Ref. \cite{Agaev:2018fvz} to explore the suggestion about a possible exotic nature of this particle. The obtained results on the mass and width of this particle  in this study do not contradict  the  experimental data, however,  more precise experimental studies were suggested.

The thermal behavior of $K_{0}^{*}(700)$, however, was investigated in few studies. Thus, in Ref. \cite{Gao:2019idb}, the authors studied the thermal properties of the lowest multiplet of the QCD light-flavor scalar resonances, including $K_{0}^{*}(700)$  state at a finite temperature in the framework of the unitarized U(3) chiral perturbation theory. They found that the mass of this resonance decreases when increasing the temperature, considerably. The thermal behavior of the $K_{0}^{*}(700)$ meson was also studied in Ref. \cite{Giacosa:2018vbw} by using an effective hadronic model. 

In the present study, we investigate the thermal behavior of the light scalar strange $K_{0}^{*}(700)$ meson. In particular, we discuss the behavior of the mass and decay constant of this state with respect to the temperature by including the hot medium effects by the fermionic and gluonic parts of the energy momentum tensor as well as the temperature dependent continuum threshold and  quark- gluon condensates. To this end, we use thermal QCD sum rule formalism. We then set $ T \rightarrow 0$ to find the value of the mass and decay constant in a vacuum. The  decay constant is one of the main input parameters to investigate the electromagnetic properties as well as possible weak and strong decays of $K_{0}^{*}(700)$, which may be in the agenda to get further information on the nature and structure of this state. The QCD sum rule approach is one of the powerful and applicable techniques to calculate the hadronid parameters \cite{Shifman}. This method was then extended to include the properties of the hadrons at finite temperature \cite{Bochkarev,Adami,Hatsuda} considering that the operator product expansion (OPE) and other assumptions of the method  remain unchanged but  the quark,  gluon and mixed condensates are changed by their thermal versions. The required extra O(3) invariance brings some additional operators with the same dimensions as the vacuum condensates in thermal sum rules.
The thermal QCD sum rules were applied to investigate many properties of the standard hadrons and exotics ( see for instance \cite{Gubler,Yazici:2015tqa,VeliVeliev:2018eaw,Veliev:2008zi} and references therein).

This work is organized in the following way: In sec. \ref{sec:Mass} we derive two-point thermal QCD sum rules for the mass and coupling constant of the $K_{0}^{*}(700)$ meson. In sec. III we perform numerical computations and discuss the thermal behaviors of the mass and decay constant of $K_{0}^{*}(700)$  with respect to temperature. In this section, we also extract values of  $m_{K_{0}^{*}}$ and $f_{K_{0}^{*}}$ in vacuum. Section \ref{sec:Conc} is reserved for our  concluding remarks.


\section{Thermal sum rules for physical quantities}

\label{sec:Mass}

This section is devoted to the calculations of the  mass and decay constant of the light strange scalar $K_{0}^{*}(700)$ meson in the
context of the QCD sum rule at a finite temperature. To this end,  we start from the following temperature-dependent two-point correlation
function:
\begin{equation}
\Pi (p,T)=i\int d^{4}xe^{ip\cdot x}\langle \mathcal{T}\{J^{K_{0}^{*}}(x)J^{K_{0}^{*}%
\dag }(0)\}\rangle_T ,  \label{eq:CorrF1}
\end{equation}%
where $J^{K_{0}^{*}}(x)$ is the interpolating field or current of the  $K_{0}^{*}(700)$ meson, $\mathcal{T}$ represents the time ordering operator and $T$ 
stands for the temperature. The  average of any operator $O$ in the medium with thermal equilibrium is written as
\begin{equation}  \label{eqn2}
\langle O\rangle_T=Tr (e^{-\beta H}O)/Tr (e^{-\beta H}), \\
\end{equation}
with $H$ being the QCD Hamiltonian and $\beta=1/T$. In the quark-antiquark picture, the scalar  current $J^{K_{0}^{*}}(x)$  is expressed by
\begin{eqnarray}
&&J^{K_{0}^{*}}(x)=d^{i}(x) \overline{s}^{i}(x) ,
\label{eq:CDiq}
\end{eqnarray}%
where $d$ and $s$ are light quarks and $i$ is the color index.

According to the general aspect of the QCD sum rules formalism, the above correlation function can be calculated in two different ways called physical and OPE representations. In order to derive QCD sum rules for the physical quantities under study,  first we evaluate the correlation function in  the hadronic language including the  parameters of hadron like its mass and decay constant. Then we calculate the same function in terms of QCD parameters and match the two representations to get the desired sum rules. Applying the Borel transformation and continuum subtraction procedures enhance the ground state  contribution and suppress the contributions of the unwanted higher states and continuum. By saturating the correlation function in Eq.\ (\ref{eq:CorrF1}) with a complete set of the $K_{0}^{*}(700)$ state and performing
an integration over $x$, we get
\begin{equation*}
\Pi^{\mathrm{Phys}}(p,T)=\frac{\langle T|J^{K_{0}^{*}}|K_{0}^{*}(p)\rangle
\langle K_{0}^{*}(p)|J^{K_{0}^{*}\dag }|T\rangle }{m_{K_{0}^{*}}^{2}(T)-p^{2}}+ \ldots,
\end{equation*}
in the zero width limit with $m_{K_{0}^{*}}(T)$ being the temperature-dependent mass of $K_{0}^{*}(700)$. Here, $ \langle T| $ represents the ground state of the medium at finite temperature and  the dots indicate contributions to the correlation function arising from the higher states and continuum. The temperature-dependent decay  constant $f_{K_{0}^{*}}(T)$ is defined using the matrix element
\begin{equation}
\langle T|J^{K_{0}^{*}}|K_{0}^{*}(q)\rangle =f_{K_{0}^{*}}(T)m_{K_{0}^{*}}(T).
\label{eq:Res}
\end{equation}
 Then in terms of $m_{K_{0}^{*}}(T)$ and $f_{K_{0}^{*}}(T)$, the correlator in the zero width limit 
is expressed as
\begin{equation}
\Pi ^{\mathrm{Phys}}(p,T)=\frac{m_{{K_{0}^{*}}}^{2}(T)f_{{K_{0}^{*}}}^{2}(T)}{m_{K_{0}^{*}}^{2}-p^{2}%
}+\ldots . \label{eq:CorM}
\end{equation}%

The Borel transformation with respect to $p^{2}$ applied to $\Pi ^{\mathrm{Phys}}(p,T)$ leads to the final form of the physical side:

\begin{eqnarray}
&&\mathcal{B}_{p^{2}}\Pi ^{\mathrm{Phys}
}(p,T)=m_{{K_{0}^{*}}}^{2}(T)f_{{K_{0}^{*}}}^{2}(T)e^{-m_{{K_{0}^{*}}}^{2}(T)/M^{2}},  \label{eq:CorBor}
\end{eqnarray}
where $ M^{2} $ is the Borel parameter to be fixed in next section. 

The OPE side of the correlation function, $\Pi ^{\mathrm{OPE}}(p,T)$, has to be determined in terms of the parameters of the quarks and gluons like quark masses, quark-gluon condensates, etc. For this aim, we insert the interpolating current presented in Eq. (\ref{eq:CDiq}) into Eq. (\ref{eq:CorrF1}), and contract the same quark fields using the Wick theorem. As a result, we get 
\begin{eqnarray}
\Pi ^{\mathrm{OPE}}(p,T)=-{i}\int d^{4}xe^{ip\cdot
x}\left\langle  \mathrm{Tr}\left[ {S}
_{d}^{ij}(x)S_{s}^{ji}(-x)\right]  \right\rangle _{T}.
\label{eq:CorrF2}
\end{eqnarray}%
The quark propagator in vacuum is given in terms of the parameters of the quarks and gluons \cite{Reinders,Wang:2009ry}.  At a nonzero temperature, the failure of the Lorentz invariance by the 
chosen reference frame and emergence of the extra
$O(3)$-symmetry some new kinds of operators appear in the OPE. In order to restore the Lorentz invariance, four-velocity vector  of the medium, $ u ^{\mu } $ is introduced. Using the four-velocity vector and quark/gluon fields, one can construct  new four dimensional  operators  like $  \langle u\Theta ^{f}u\rangle $, where $\Theta _{\mu \nu }^{f}$ is the fermionic part of the energy-momentum tensor (for more information see for instance Refs. \cite{Bochkarev,Adami,Hatsuda}).  Thus the  light-quark propagator in a hot medium\cite{Mallik:1997pq} can be written as
\begin{eqnarray}
S_{q}^{ij}(x) &=&i\frac{\slashed
x}{2\pi^{2}x^{4}}\delta_{ij}-\frac{
m_{q}}{4\pi^{2}x^{2}}\delta_{ij}  \notag \\
&-&\frac{\langle \bar{q}q\rangle }{12}\delta_{ij}-\frac{x^{2}}{192}%
m_{0}^{2}\langle \bar{q}q\rangle \Big[1-i\frac{m_{q}}{6}\slashed x \Big]%
\delta _{ij}  \notag \\
&+&\frac{i}{3}\Big[\slashed x \Big(\frac{m_{q}}{16}\langle
\bar{q}q\rangle
-\frac{1}{12}\langle u\Theta ^{f}u\rangle \Big)  \notag \\
&+&\frac{1}{3}\Big(u\cdot x\Big)\slashed u \langle u\Theta ^{f}u\rangle %
\Big]\delta _{ij}  \notag \\
&-&\frac{ig_{s}\lambda _{ij}^{A}}{32\pi ^{2}x^{2}}G_{A}^{\mu \nu }
\Big(\slashed x \sigma _{\mu \nu }+\sigma _{\mu \nu }\slashed
x\Big), \label{lightquarkpropagator}
\end{eqnarray}%
where $m_{q}$ represents the light $ s $ or $ d $ quark mass,  $\langle \bar{q}q\rangle $ stands for the  light quark condensate in the hot medium and $G_{A}^{\mu \nu }$ shows the external gluon field. In Eq.\ (\ref{lightquarkpropagator}) $i,\,j$ are color indices and $\lambda
_{A}^{ij}$ are Gell-Mann matrices with $A$  runs from  $1$ to $ 8 $. 

The correlation function $\Pi^{\mathrm{OPE}}(p,T)$ can be
expressed in terms of two parts:  perturbative and non-perturbative. The perturbative part is written in terms of a dispersion integral. Hence,
\begin{eqnarray}
\Pi ^{\mathrm{OPE}}(p,T) &=&\int_{(m_{d}+m_{s})^{2}}^{s_{0}(T) }\frac{\rho (s)}{s-p^{2}}ds  \notag \\
&&+\Pi ^{n.pert}(p,T),
\end{eqnarray}%
where $ s_{0}(T) $ is the temperature-dependent continuum threshold and $ \rho (s) $ is the spectral density, which is obtained using the imaginary part of the correlation function.   For the channel under discussion, $ \rho (s) $ is obtained as
\begin{eqnarray}
\rho (s) &=&\frac{3(2m_d m_s-s)}{8\pi^2}.
\end{eqnarray}
The function $\Pi^{\mathrm{OPE}}(p,T)$ in Borel scheme reads
\begin{eqnarray}
\mathcal{B}_{p^{2}}\Pi ^{\mathrm{OPE}}(p,T) &=&\int_{(m_{d}+m_{s})^{2}}^{s_{0}(T)}\rho (s)e^{-s/M^{2}}ds  \notag \\
&&+\mathcal{B}_{p^{2}}\Pi ^{n.pert}(p,T),
\end{eqnarray}%
where the  non-perturbative part in QCD is obtained as:
\begin{eqnarray}
&&\mathcal{B}_{p^{2}}\Pi ^{n.pert}(p,T) =\frac{\langle \bar{d}d\rangle (2m_s+m_d)}{2} \notag \\
&&+\frac{\langle \bar{s}s\rangle (2m_d+m_s)}{2}+\frac{1}{8}\langle\alpha_{s}\frac{G^2}{\pi}%
\rangle -\frac{g_s^2\langle u \Theta^g u\rangle}{24\pi^2 }  \notag \\
&&-\frac{4\langle u \Theta^f u\rangle}{3} .
\end{eqnarray}%

The QCD sum rules for the spectroscopic parameters are
extracted following the  matching of the functions  $\mathcal{B}_{p^{2}}\Pi ^{\mathrm{Phys}}(p,T)$ and $\mathcal{B}_{p^{2}}\Pi ^{\mathrm{%
OPE}}(p,T)$. 
In this study,  we take into
account the quark, gluon as well as their mixed condensates up to dimension ten. However,  contributions of the   operators with mass dimensions five and  higher  are obtained to be zero.

  The following expression is considered to
write down the gluon condensate entering the calculations according to the
gluonic term of the energy-momentum tensor, $\Theta _{\lambda \sigma }^{g}$ (for
details see for instance Ref. \cite{Mallik:1997pq}):
\begin{eqnarray}
&&\langle Tr^{c}G_{\alpha \beta }G_{\mu \nu }\rangle =\frac{1}{24}(g_{\alpha
\mu }g_{\beta \nu }-g_{\alpha \nu }g_{\beta \mu })\langle G_{\lambda \sigma
}^{a}G^{a\lambda \sigma }\rangle   \notag  \label{TrGG} \\
&&+\frac{1}{6}\Big[g_{\alpha \mu }g_{\beta \nu }-g_{\alpha \nu }g_{\beta \mu
}-2(u_{\alpha }u_{\mu }g_{\beta \nu }-u_{\alpha }u_{\nu }g_{\beta \mu }
\notag \\
&&-u_{\beta }u_{\mu }g_{\alpha \nu }+u_{\beta }u_{\nu }g_{\alpha \mu })\Big]%
\langle u^{\lambda }{\Theta }_{\lambda \sigma }^{g}u^{\sigma }\rangle .
\end{eqnarray}

The mass sum rule for the $K_{0}^{*}(700)$ meson is obtained as
\begin{eqnarray}
m_{K_{0}^{*}}^{2}(T)=\frac{\int_{(m_{d}+m_{s})^{2}}^{s_{0}(T)}dss\rho
(s)e^{-s/M^{2}}+\widetilde{\Pi}^{n.pert}}{\int_{(m_{d}+m_{s})^{2}}^{s_{0}(T)}ds\rho
(s)e^{-s/M^{2}}+\mathcal{B}\Pi ^{n.pert}}
\end{eqnarray}%
where
\begin{eqnarray}
\widetilde{\Pi}^{n.pert}=- \frac{d}{d(1/M^2)}\mathcal{B}\Pi^{n.pert},
\end{eqnarray}
with $ \mathcal{B}\Pi ^{n.pert}= \mathcal{B}_{p^{2}}\Pi ^{n.pert}(p,T)$.

Finally,  the decay constant   $f_{K_{0}^{*}}(T)$ is calculated  from the sum rule
\begin{eqnarray}
f_{K_{0}^{*}}^{2}(T)&=&\frac{1}{m_{K_{0}^{*}}^{2}(T)}
\Bigg\lbrace\int_{(m_{d}+m_{s})^{2}}^{s_{0}(T)}ds\rho
(s)e^{(m_{K_{0}^{*}}^{2}(T)-s)/M^{2}} \nonumber  \\
&+&\mathcal{B}\Pi ^{n.pert}\Bigg\rbrace.
\end{eqnarray}%

\section{Numerical Analyses}

\label{sec:Num}

The expressions for the thermal  QCD sum rules for the mass and decay constant of the $K_{0}^{*}(700)$ meson include  various parameters such as temperature-dependent  quark and  gluon condensates, the gluonic and fermionic parts of the energy momentum tensor, temperature-dependent continuum threshold as well as the Borel parameter. 
\begin{table}[tbp]
\begin{tabular}{|c|c|}
\hline\hline
Parameters & Values \\ \hline\hline
$\langle 0|\bar{q}q|0\rangle  $ & $(-0.241\pm 0.01)^3 ~\mathrm{GeV}^3$ \cite%
{Shifman,Reinders} \\
$\langle 0|\frac{\alpha_sG^2}{\pi}|0\rangle$ & $(0.012\pm0.004)~\mathrm{GeV}^4$
\cite{Shifman,Reinders} \\
$ m_s$ & $93^{+11}_{-5}  ~\mathrm{MeV}$ \cite{Tanabashi:2018oca} \\
$ m_d$ & $4.67^{+0.48}_{-0.17}  ~\mathrm{MeV}$ \cite{Tanabashi:2018oca} \\
\hline\hline
\end{tabular}%
\caption{Input parameters.}
\label{tab:Param}
\end{table}

 For the temperature-dependent quark condensate, we use the following  fit function
extracted in Refs. \cite{Azizi:2016ddw,Ayala}: 
\begin{equation}
\langle \bar{q}q\rangle =\frac{\langle 0|\bar{q}q|0\rangle }{%
1+e^{18.10042(1.84692[\frac{1}{\mathrm{GeV}^{2}}]T^{2}+4.99216[\frac{1}{%
\mathrm{GeV}}]T-1)}}.
\end{equation}%
This parametrization, which is  reliable up to the  critic temperature $T_{c}=197~\mathrm{MeV}$, has been obtained by fitting it to  the lattice QCD results borrowed from Refs. \cite{Bazavov,Cheng1}. Here,  $\langle 0|\bar{q}q|0\rangle $ denotes the vacuum light-quark condensate, whose value is presented in  table \ref{tab:Param}.

For the thermal gluon condensate, we use the fit function, which has been  extracted using both the QCD sum rule   and lattice QCD results in Refs.
\cite{Azizi:2016ddw,Ayala2}:
\begin{eqnarray}
\langle G^{2}\rangle  &=&\langle 0|G^{2}|0\rangle \Bigg[1-1.65\Big(\frac{T}{%
T_{c}}\Big)^{8.735}  \notag \\
&+&0.04967\Big(\frac{T}{T_{c}}\Big)^{0.7211}\Bigg],
 \label{G2TLattice} 
\end{eqnarray}%
where $\langle 0|G^{2}|0\rangle $ is the vacuum gluon condensate,  whose value is presented in  table \ref{tab:Param}, as well. In table \ref{tab:Param}, we also present the values of the light quark masses used in the calculations.

Finally, for the fermionic and gluonic contributions of the energy-momentum tensor we make use of 
the fit function extracted in Ref. \cite{Azizi:2016ddw} by using
the lattice QCD results on the thermal behavior of the energy-momentum tensor from Ref.  \cite{Cheng:2007jq}:
\begin{eqnarray}
&&\langle \Theta _{00}^{g}\rangle =\langle \Theta _{00}^{f}\rangle
=\dfrac{1}{2}\langle \Theta _{00}\rangle \nonumber\\
&=&T^{4}e^{[113.867(\frac{1}{GeV^{2}})T^{2}-12.190(\frac{1}{GeV})T]} -10.141(\frac{1}{GeV})T^{5}.\nonumber\\
\label{tetamumu}
\end{eqnarray}

The temperature-dependent continuum threshold in light systems are taken as $s_0(T)\simeq s_0 \frac{\langle \bar{q}q\rangle}{\langle 0|\bar{q}q|0\rangle} $. The exact relation is found by imposing  the conditions that $ s_0(T) $ reduces to the vacuum threshold at a zero temperature, reflects the temperature behavior of quark condensates and the pole dominance as well as OPE convergence at all temperatures are satisfied.  These lead to the expression, 
\begin{eqnarray}  \label{G2TLattice}
s_0(T)&=&s_0\left[1-0.2 \Big(\frac{T}{T_C}\Big)^{4}-0.7 \Big(\frac{T}{T_C}\Big)^{12}\right],
\end{eqnarray}
where $s_0$ is the vacuum  threshold. The continuum threshold in a vacuum  is determined using the requirements of the method such as the stability of the results with respect to its variations as well as considering  the energy of the first excited meson in the  $K_{0}^{*}(700)$ channel. For $s_0$, we choose the range
\begin{equation}
1.05~\mathrm{GeV}^2\leq s_0\leq 1.25~\mathrm{GeV}^2,  \label{eq:19}
\end{equation}
where the mass and decay constant of $K_{0}^{*}(700)$ meson  show good stability with respect to its changes.

Based on  the prescriptions  of the QCD sum rule approach, the mass and decay constant of  the $K_{0}^{*}(700)$ meson  should also show mild variations with respect to $M^2$. The working window for the Borel parameter $M^2$ is acquired by demanding that the higher state and continuum contributions are small and the contributions coming from the higher dimensional operators are suppressed. In other words, to determine the window for Borel parameter, we impose the conditions of the pole dominance and OPE convergence at a zero temperature. Therefore, in the present study, we fix the following working window for $M^2$:
\begin{equation}
0.8~\mathrm{GeV}^2\leq M^2\leq 1.4~\mathrm{GeV}^2.  \label{eq:19â}
\end{equation}

\begin{widetext}

\begin{figure}[ht]
\begin{center}
\subfigure[]{\includegraphics[width=8cm]{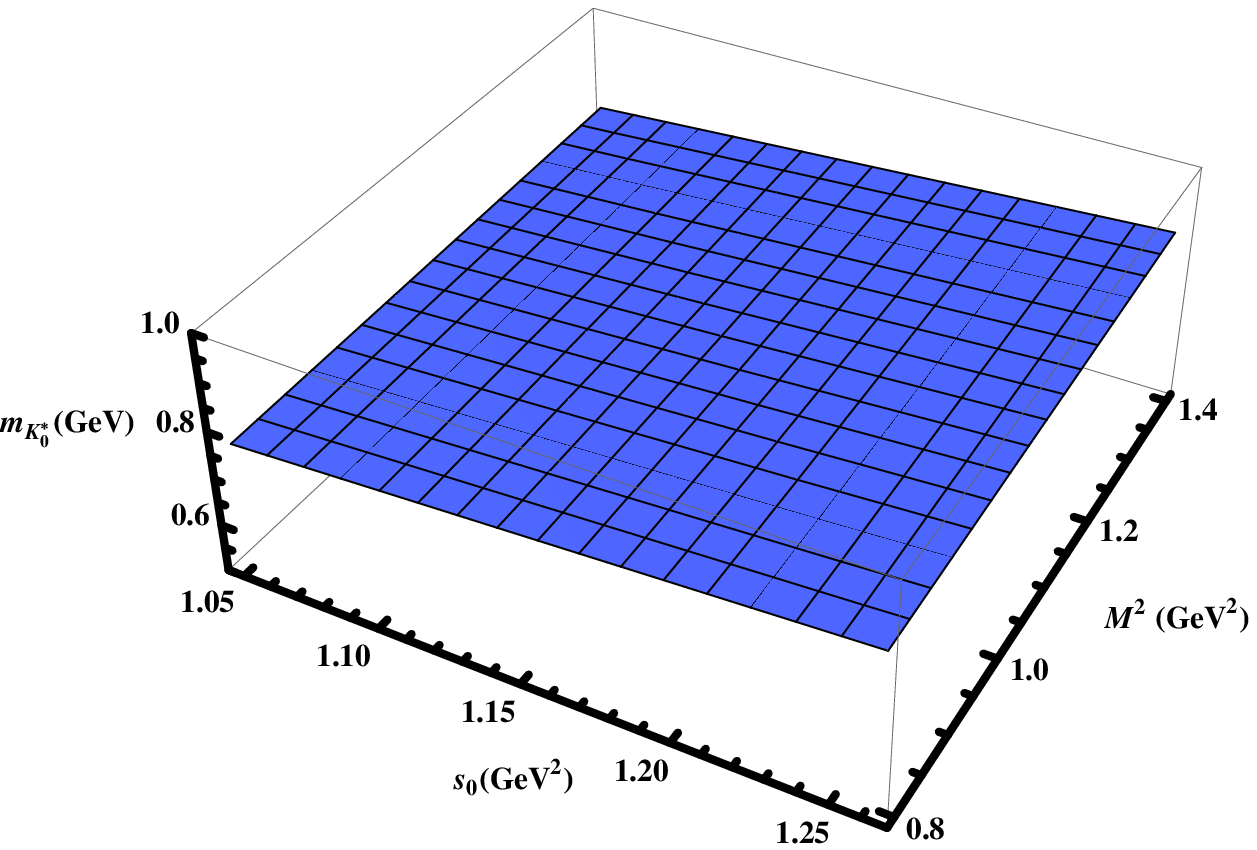}}
\subfigure[]{\includegraphics[width=8cm]{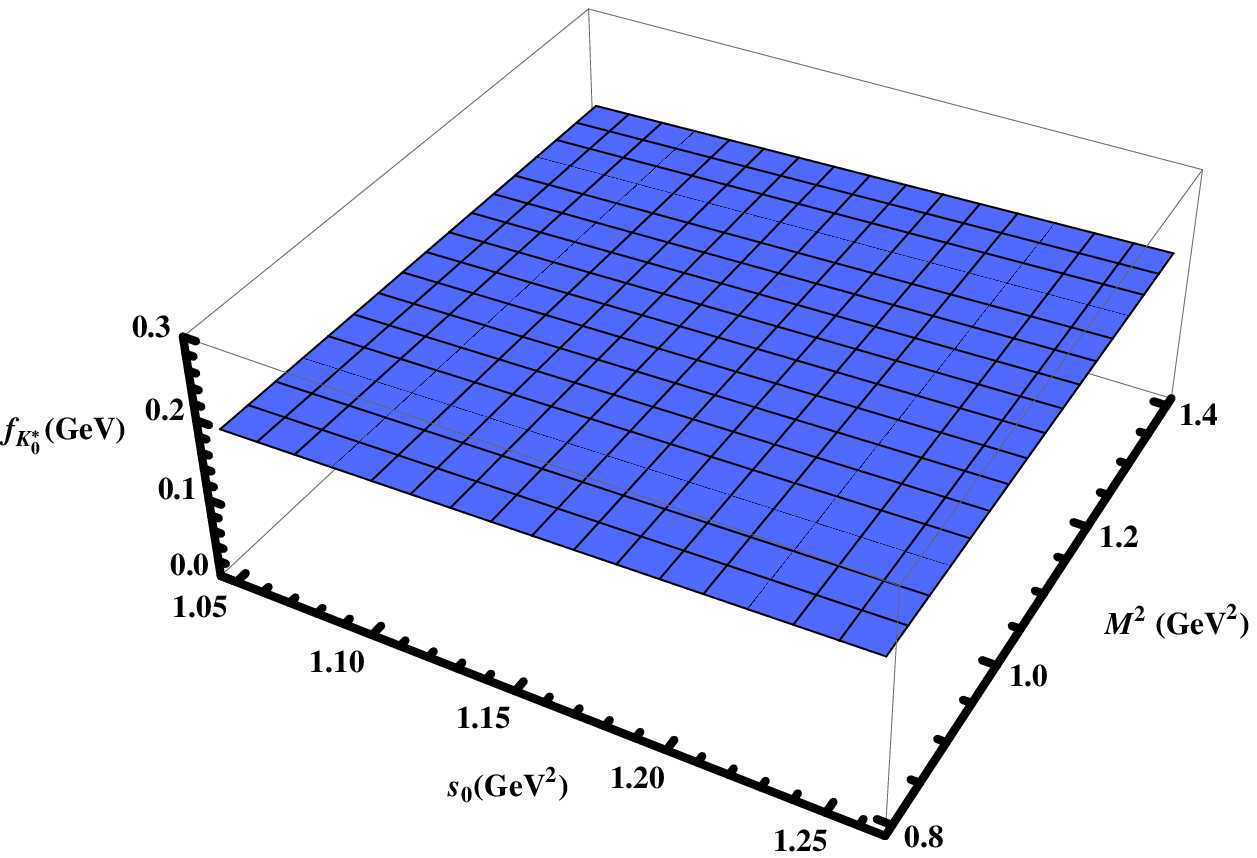}}
\end{center}
\caption{(a) Dependence of the vacuum mass of the  $K_{0}^{*}(700)$ state on $M^2$ and $s_0$. (b)  The same
as (a) but for the decay constant $f_{K_{0}^{*}}$.} \label{fig1}
\end{figure}
\end{widetext}
\begin{table}[tbp]
\begin{tabular}{|c|c|c|}
\hline\hline
& $m_{K_{0}^{*}}(\mathrm{MeV})$ & $f_{K_{0}^{*}} (\mathrm{MeV})$
\\ \hline\hline
Present work & $820\pm10$ & $191\pm 4$ \\ \hline
Experiment \cite{Tanabashi:2018oca} & $824\pm30$ & - \\ \hline
\end{tabular}%
\caption{Vacuum mass and decay constant values for  the $K_{0}^{*}(700)$ meson.}
\label{tab:Results1}
\end{table}
The 3-D mass and decay constant graphics for the $K_{0}^{*}(700)$ meson  in a vacuum are presented in figure \ref{fig1}. From this figure, we see that the mass and decay constant show mild variations with respect to the changes in  $M^2$ and $ s_0 $ and satisfy the requirements of the method used. Extracted from the analyses, the vacuum results for the mass and decay constant of the meson $K_{0}^{*}(700)$ are depicted in table \ref{tab:Results1}. The world average for the Breit-Wigner mass from the experiment presented in PDG is also shown in the same table. We observe that our result is in a nice consistency with the experimental data. The error presented in our prediction is small in comparison with the experimental uncertainty. The errors in our results for the mass and decay constant are due to the uncertainties in calculations of the working windows for the auxiliary parameters as well as those related to other input parameters. These errors are  small  compared to the limits allowed by the sum rule calculations. Nevertheless, roughly $ 60\% $ of the presented errors belong to the variations of the results with respect to the auxiliary parameters $M^2$ and $ s_0 $ and roughly $ 40\% $ are coming from the uncertainties of the input parameters, that is, the quark and gluon condensates as well as the strange quark mass. 

\begin{widetext}

\begin{figure}[ht]
\begin{center}
\subfigure[]{\includegraphics[width=8cm]{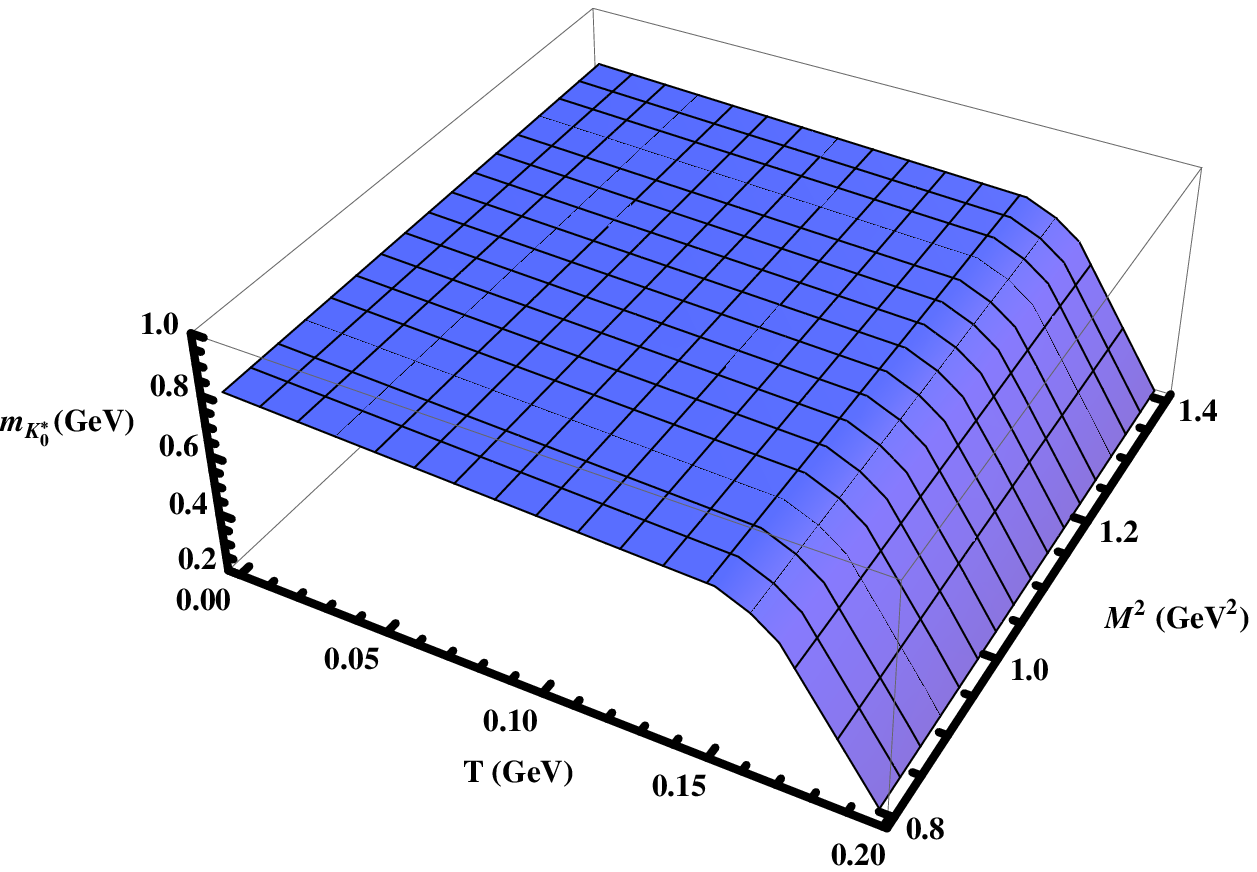}}
\subfigure[]{\includegraphics[width=8cm]{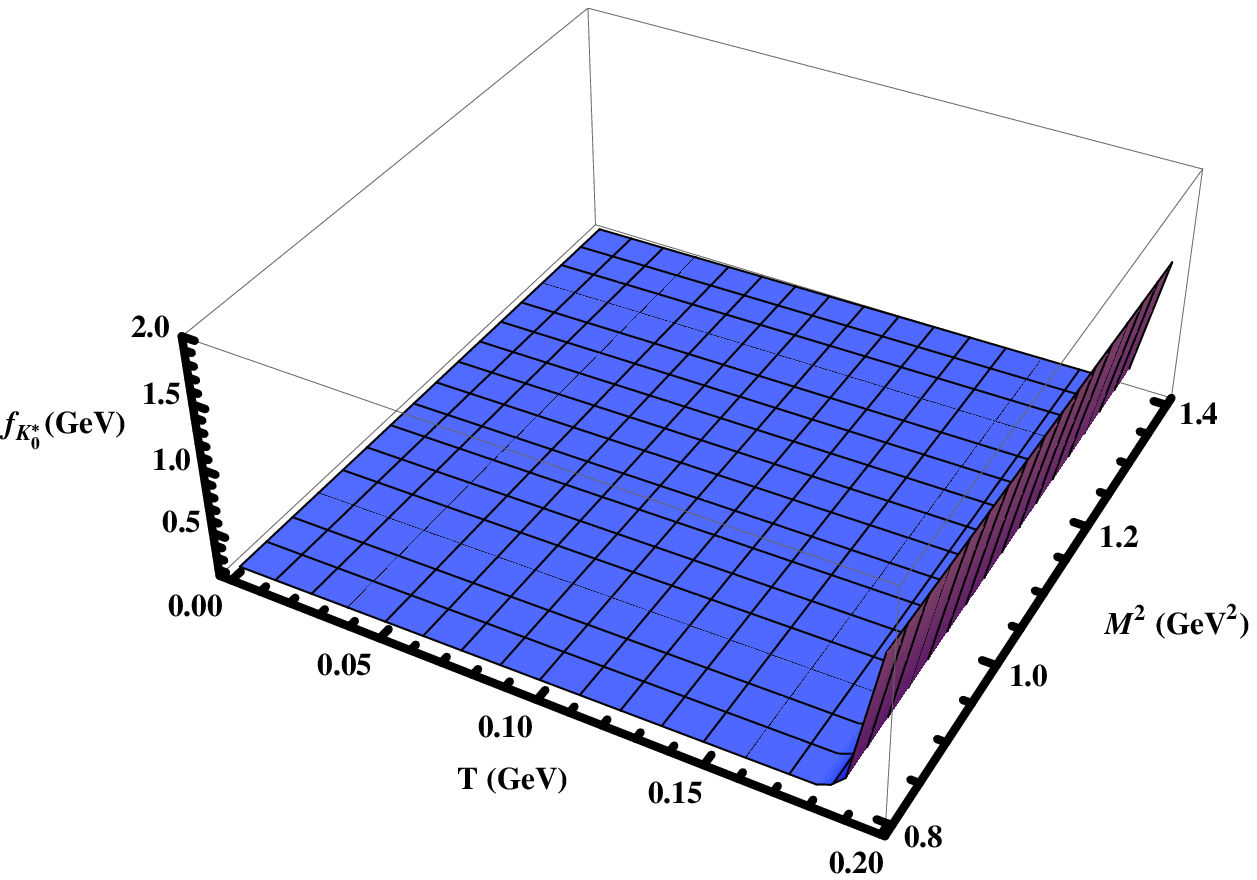}}
\end{center}
\caption{(a) Dependence of the mass of the  $K_{0}^{*}(700)$ state on $M^2$ and $T$ at an average value of the continuum threshold. (b)  The same
as (a) but for the decay constant $f_{K_{0}^{*}}$.} \label{fig8}
\end{figure}

\end{widetext}

Now, we  discuss the behavior of the mass and decay constant of the light scalar strange $K_{0}^{*}(700)$  meson with respect to the temperature. To this end we depict the 3-D graphics (see figure \ref{fig8}) showing the variations of the mass and decay constant with respect to the temperature as well as $M^2$
at average value of the continuum threshold. We see that the mass of $K_{0}^{*}(700)$ remains unchanged up to $T \simeq 0.6 ~ T_c$  but it starts to diminish after this point and approaches to zero near to the critical temperature reffering to the melting of the meson at $ T_c $. The decay constant of $K_{0}^{*}(700)$ is mild up to $T \simeq 0.85 ~ T_c$, but starts to rapidly grow after this point.

\begin{figure}[ht]
\begin{center}
\subfigure[]{\includegraphics[width=8cm]{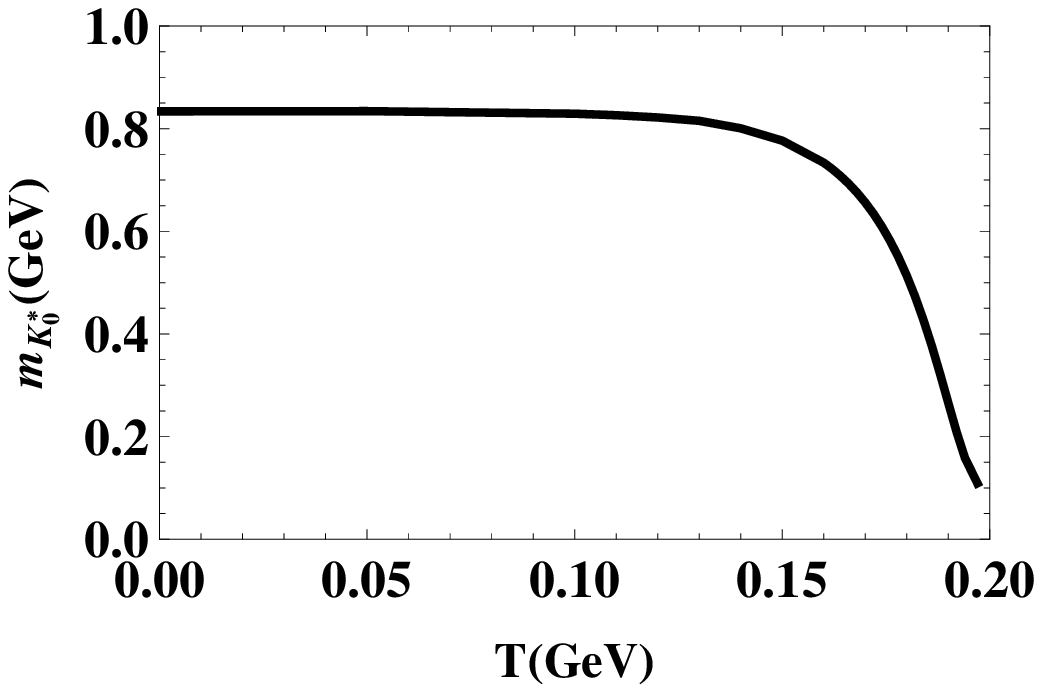}}
\subfigure[]{\includegraphics[width=8cm]{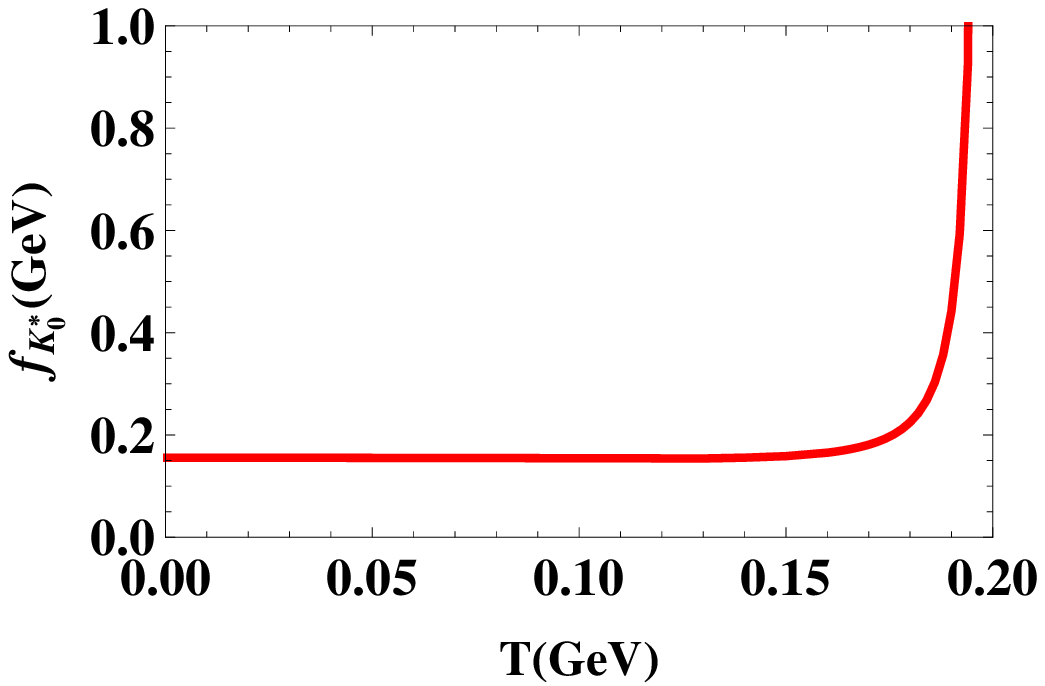}}
\subfigure[]{\includegraphics[width=8cm]{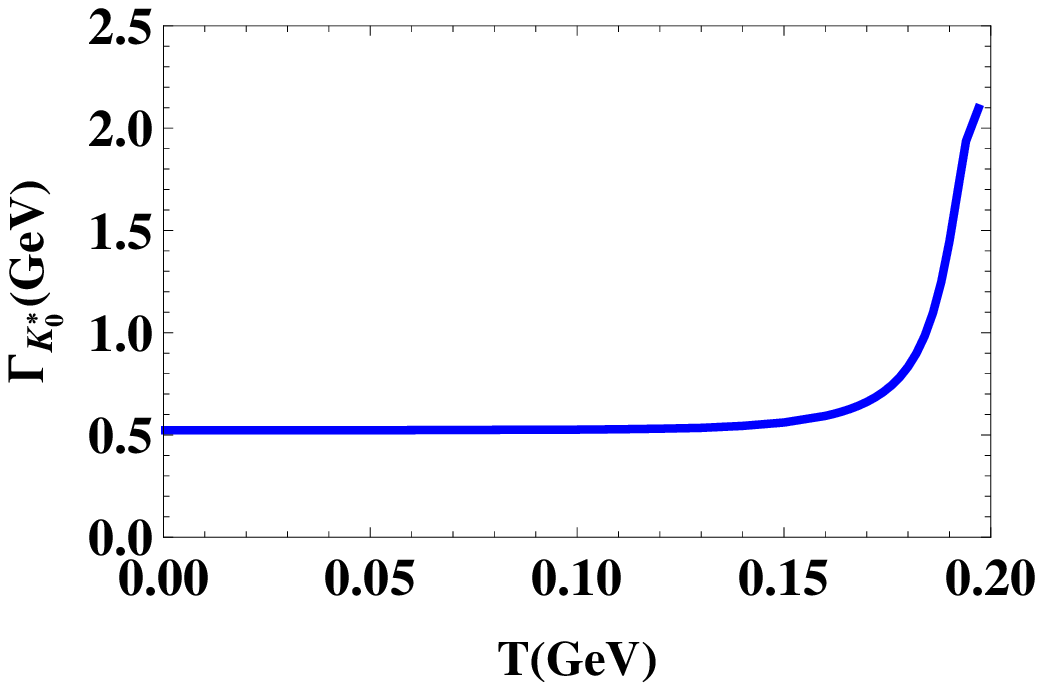}}
\end{center}
\caption{(a) Dependence of the mass of the  $K_{0}^{*}(700)$ state on  $T$ for the case of finite width and at average values of the  auxiliary parameters. (b)  The same
as (a) but for the decay constant $f_{K_{0}^{*}}$. (c) The same
as (a) but for the width $\Gamma_{K_{0}^{*}}$. }\label{figgg}
\end{figure}

At the end of this section, we  discuss the effects of considering the  finite width in the sum rules on the vacuum values of the parameters under consideration and their thermal behavior. 
At finite width, the  main sum rule obtained by matching the physical and OPE  sides in Borel scheme  takes the form (see for instance Ref. \cite{Azizi:2010zza} for more details):
\begin{eqnarray}
\label{finitewidth}
&&\frac{2}{\pi}m_{K_{0}^{*}}^{3}(T)f_{{K_{0}^{*}}}^{2}(T) \Gamma_{K_{0}^{*}}(T)\times \nonumber\\&&\int^{\infty}_0\frac{e^{-s/M^2}}{[s-m_{K_{0}^{*}}^{2}(T) ]^{2}+m_{K_{0}^{*}}^{2}(T) \Gamma_{K_{0}^{*}}^{2}(T)}=\mathcal{B}_{p^{2}}\Pi ^{\mathrm{OPE}}(p,T) .\nonumber\\
\end{eqnarray}%
To find the temperature-dependent mass, width and decay constant in this case we need two more equations, which are found via successive application of the operator $ \frac{d}{d(-\frac{1}{M^{2}})} $  to both sides of the above equation. By simultaneous solving of the  resultant three equations one finds the three unknowns: $ m_{K_{0}^{*}} (T)$, $f_{{K_{0}^{*}}} (T) $ and $ \Gamma_{K_{0}^{*}}(T) $.  In figure \ref{figgg}, we present the variations of these three quantities with respect to temperature for the case of  finite width. 
From this figure we see that although the values of mass and decay constant are considerably  shifted compared to the case of a zero width approximation, the behaviors of mass and decay constant remain unchanged, i.e., the mass rapidly falls and the decay constant rapidly grows near  the critical temperature.  As  is seen, the width of $K_{0}^{*}(700)$  remains unchanged up to roughly $ T=150~\mathrm{MeV} $, after which it starts to grow up to the critical temperature, considerably. 
At the $ T\rightarrow 0  $ limit we obtain the values of the quantities under consideration as 
\begin{eqnarray}
 m_{K_{0}^{*}} (0)=834\pm 10~\mathrm{MeV} , \nonumber\\
  f_{K_{0}^{*}} (0)=156\pm 3~\mathrm{MeV} , \nonumber\\
  \Gamma_{K_{0}^{*}} (0)=524\pm 8~\mathrm{MeV} , 
\end{eqnarray}
where show considerable differences compared to the values of the mass and decay constant given in table II at zero width approximation. The central value of the mass is shifted with $+14 ~\mathrm{MeV}  $, while this amount in decay constant is $ -35~\mathrm{MeV}  $.  The values of the mass and width, within the errors, are consistent with the  experimental Breit-Wigner mass and width values reported by PDG.

\section{Concluding Remarks}
\label{sec:Conc}
Despite a lot of experimental and theoretical effort, the nature and structure of the light scalar mesons remain unclear and their parameters suffer from large uncertainties. In the case of $K_{0}^{*}(700)$  the situation is even worse: The label $(700)$ on its name differs with its mass considerably and its parameters include large uncertainties in the experiment. To clarify the situation with this lightest scalar strange meson, we calculated the mass and decay constant of this state in the framework of thermal QCD. We obtained a vacuum  mass at a zero width limit in accord with the world average  Breit-Wigner mass from the experiment presented in PDG. In the optimal working windows of the auxiliary parameters our results encompass small uncertainties compared with the experimental data. Our results on the mass and decay constant may help experimental groups to more clarify the situation. The decay constant obtained in vacuum can be used as one of the main input parameters to investigate the electromagnetic properties as well as the weak and strong interactions of $K_{0}^{*}(700)$ meson with other known particles.

We discussed the thermal behaviors of the mass and decay constant and observed that the mass and decay constant remain constant up to $T \simeq 0.6 ~ T_c$ and $T \simeq 0.85 ~ T_c$, respectively. After these points the mass starts to fall and the decay constant grows rapidly:  The mass approaches to zero at the critical temperature, referring to the melting of $K_{0}^{*}(700)$ meson and its decay constant grows substantially at $ T_c $.

By considering the finite width in the calculations, we observed considerable shifts in the values of the mass and decay constant, although their thermal behavior was not changed.  The calculations at a finite width show that the width of $K_{0}^{*}(700)$  remains unchanged up to roughly $ T=150~\mathrm{MeV} $, after which it starts to grow, considerably. At finite width, we also calculated the vacuum mass and width of $K_{0}^{*}(700)$, whose values are consistent with the experimental Breit-Wigner mass and width values within the errors. 

 With the progress made in the   construction of the future in-medium experiments  such as Japan proton accelerator research complex (JPARC), compressed baryonic matter (CBM) and   anti-proton annihilation  Darmstadt (PANDA) at GSI Germany, as well as  nuclotron-based ion collider facility (NICA) at  Dubna Russia,  it will be possible to test the behaviors of hadrons at a finite temperature and density. Comparison of the future data with the phenomenological predictions will help us clarify the situation with the scalar mesons and  get valuable knowledge on their nature and quark-gluon organization. This will also shed light on the non-pertubative nature of QCD at a finite temperature and density. 

\begin{widetext}
\section*{Acknowledgments}
H. S. thanks Kocaeli University for the support provided under  Grant no BAP 2019/064HD.
\label{sec:e}
\end{widetext}



\begin{thebibliography}{999}


\bibitem{Jaffe} R. L. Jaffe, Phys. Rev. D {\bf 15}, 267 (1977).
\bibitem{Weinstein} J. D. Weinstein and N. Isgur, Phys. Rev. D {\bf 41}, 2236 (1990).
\bibitem{PDG}  M. Tanabashi et al. (Particle Data Group), Phys. Rev. D {\bf 98}, 030001 (2018) and 2019 update.
  \bibitem{BES-II} F.K. Guo et al., Nucl. Phys. A {\bf 773}, 78 (2006).
    \bibitem{Belle} D. Epifanov et al., Phys. Lett. B {\bf 654}, 65 (2007).
\bibitem{Cawlfield} C. Cawlfield et al., Phys. Rev. D {\bf 74}, 031108R (2006).
\bibitem{Anisovich} A. V. Anisovich and A.V. Sarantsev, Phys. Lett. B {\bf 413}, 137 (1997).
\bibitem{Delbourgo} R. Delbourgo et al., Int. J. Mod. Phys. A {\bf 13}, 657 (1998).
\bibitem{Oller} J. A. Oller and E. Oset, Phys. Rev. D {\bf 60}, 074023 (1999).
\bibitem{Shakin} C. M. Shakin and H. Wang, Phys. Rev. D {\bf 63}, 014019 (2001).
\bibitem{Scadron} M. D. Scadron et al., Nucl. Phys. A {\bf 724}, 391 (2003).
\bibitem{Bugg} D. V. Bugg, Phys. Lett. B {\bf 572}, 1 (2003).
\bibitem{Zheng} H. Q. Zheng et al., Nucl. Phys. A {\bf 733}, 235 (2004).
\bibitem{Zhou} Z. Y. Zhou and H.Q. Zheng, Nucl. Phys. A {\bf 775}, 212 (2006).
\bibitem{Link} J. M. Link et al., Phys. Lett. B {\bf 653}, 1 (2007).
\bibitem{Aubert} B. Aubert et al., Phys. Rev. D {\bf 76}, 011102R (2007).
\bibitem{Kopp} S. Kopp et al., Phys. Rev. D {\bf 63}, 092001 (2001).
\bibitem{Jamin} M. Jamin et al., Nucl. Phys. B {\bf 587}, 331 (2000).
\bibitem{Black} D. Black, Phys. Rev. D {\bf 64}, 014031 (2001).
\bibitem{Genon} S. Descotes-Genon and B. Moussallam, Eur. Phys. J. C {\bf 48}, 553 (2006).
\bibitem{Pelaez} J. R. Pelaez, A. Rodas, J. Ruiz de Elvira, Eur. Phys. J. C {\bf 77}, 91 (2017).
\bibitem{Humanic:2018brf} 
  T.~J.~Humanic,
  J.\ Phys.\ G {\bf 46}, no. 5, 055001 (2019).
\bibitem{Alford:2000mm} 
  M.~G.~Alford and R.~L.~Jaffe,
  Nucl.\ Phys.\ B {\bf 578}, 367 (2000)
  
\bibitem{Agaev:2018fvz} 
  S.~S.~Agaev, K.~Azizi and H.~Sundu,
  Phys.\ Lett.\ B {\bf 789}, 405 (2019).
  
  
\bibitem{Gao:2019idb} 
  R.~Gao, Z.~H.~Guo and J.~Y.~Pang,
  arXiv:1907.01787 [hep-ph].
\bibitem{Giacosa:2018vbw} 
  F.~Giacosa,
  arXiv:1811.00298 [hep-ph].
    
    
    
    
    
    

\bibitem{Shifman}  M. A. Shifman, A. I. Vainshtein and V. I. Zakharov,
Nucl. Phys. B \textbf{147}, 385 (1979).



\bibitem{Bochkarev} A.I. Bochkarev, M.E. Shaposhnikov, Nucl. Phys. B \textbf{268}, 220 (1986).

\bibitem{Adami} C. Adami, T. Hatsuda, I. Zahed, Phys. Rev. D \textbf{43}, 921 (1991).
\bibitem{Hatsuda} T. Hatsuda, Y. Koike, S. H. Lee, Nucl. Phys. B \textbf{394}, 221 (1993).




\bibitem{Gubler} P. Gubler, K. Morita and M. Oka, Phys. Rev. Lett. \textbf{%
107}, 092003 (2011).


  
  
  
 
  
  
\bibitem{Yazici:2015tqa} 
  E.~Yazici, H.~Sundu and E.~V.~Veliev,
  Eur.\ Phys.\ J.\ C {\bf 76}, no. 2, 89 (2016)
  
\bibitem{VeliVeliev:2018eaw} 
  E.~Veli Veliev, S.~Gunaydin and H.~Sundu,
  Eur.\ Phys.\ J.\ Plus {\bf 133}, no. 4, 139 (2018).
\bibitem{Veliev:2008zi} 
  E.~V.~Veliev and T.~M.~Aliev,
  J.\ Phys.\ G {\bf 35}, 125002 (2008). 





\bibitem{Reinders} L. J. Reinders, H. Rubinstein and S. Yazaki,
Phys. Rept. 127, 1(1985).


\bibitem{Wang:2009ry}  Z.~G.~Wang, Z.~C.~Liu and X.~H.~Zhang,
Eur.\ Phys.\ J.\ C \textbf{64}, 373 (2009).







\bibitem{Mallik:1997pq}  S.~Mallik,
Phys.\ Lett.\ B \textbf{416}, 373 (1998).





\bibitem{Ayala} A. Ayala, A. Bashir, C. A. Dominguez, E. Gutierrez, M.
Loewe, A. Raya, 
Phys. Rev. D \textbf{84}, 056004 (2011).

\bibitem{Azizi:2016ddw}  K.~Azizi and G.~Bozk{\i}r,
Eur.\ Phys.\ J.\ C \textbf{76}, no. 10, 521 (2016).


\bibitem{Cheng1} M. Cheng et al.,
Phys. Rev. D \textbf{81}, 054504 (2010).

\bibitem{Bazavov} A. Bazavov et al.,
Phys. Rev. D \textbf{80}, 014504 (2009).

\bibitem{Ayala2} A. Ayala, C. A. Dominguez, M. Loewe, Y. Zhang,
Phys. Rev. D \textbf{86}, 114036 (2012).



\bibitem{Cheng:2007jq}  M.~Cheng \textit{et al.},
Phys.\ Rev.\ D \textbf{77}, 014511 (2008). 

   

\bibitem{Tanabashi:2018oca} 
  M.~Tanabashi {\it et al.} [Particle Data Group],
  Phys.\ Rev.\ D {\bf 98}, no. 3, 030001 (2018).

\bibitem{Azizi:2010zza} 
  K.~Azizi and N.~Er,
  Phys.\ Rev.\ D {\bf 81}, 096001 (2010).

\end{thebibliography}
\end{document}